\documentstyle[aps,pra,twocolumn]{revtex}
\begin{document}
\draft

\newcommand{\pp}[1]{\phantom{#1}}
\newcommand{\be}{\begin{eqnarray}}
\newcommand{\ee}{\end{eqnarray}}
\newcommand{\ve}{\varepsilon}
\newcommand{\vs}{\varsigma}
\newcommand{\Tr}{{\,\rm Tr\,}}
\newcommand{\pol}{\frac{1}{2}}

\title{
Microscopic Foundation of Nonextensive Statistics
}
\author{Marek~Czachor$^{1,2}$ and Jan Naudts$^2$}
\address{
$^1$Katedra Fizyki Teoretycznej i Metod Matematycznych,
 Politechnika Gda\'{n}ska\\
ul. Narutowicza 11/12, 80-952 Gda\'{n}sk, Poland\\
$^2$Department of Physics, University of Antwerp, UIA, 2610
Antwerpen, Belgium
}
\maketitle

\begin{abstract}

\noindent
Combination of the Liouville equation with the $q$-averaged energy
$U_q=\langle H\rangle_q$ leads to a microscopic framework for
nonextensive $q$-thermodynamics. The resulting von Neumann equation
is nonlinear: $i\dot\rho=[H,\rho^q]$. In spite of its nonlinearity
the dynamics is consistent with linear quantum mechanics of pure
states. The free energy $F_q=U_q-TS_q$ is a stability function for
the dynamics. This implies that $q$-equilibrium states are
dynamically stable. The (microscopic) evolution of $\rho$ is
reversible for any $q$, but for $q\neq 1$ the corresponding
macroscopic dynamics is irreversible.
\end{abstract}

\bigskip
Standard thermodynamics is based on the Gibbs-Shannon-von
Neumann entropy $S_1=-k_B\Tr(\rho\ln\rho)$ and the internal energy
$U_1=\Tr\rho H$. The equilibrium density matrix $\rho_0$
minimizes the free energy $F_1=U_1-TS_1$. 
A microscopic foundation for thermodynamics is based on the
von Neumann equation (vNE)
\be
i\dot \rho=[H,\rho].\label{vN}
\ee
Equilibrium states are stable fixpoints of this dynamics.

It is perhaps not so widely known that the vNE can be regarded as a
classical Hamiltonian system with Hamiltonian {\it function\/}
$U_1$. In this context $F_1$ is a stability function for the
underlying Hamiltonian Lie-Poisson dynamics.  We shall elaborate on
these points later but first we want to pose the following problem.
It is known that there exist physical systems that are naturally
described by a nonextensive thermodynamics \cite{Tsallis}. The
structure of this theory is analogous to the ordinary one, with
one exception: Instead of $U_1$, $S_1$ and $F_1$ one takes their
$q\neq 1$ generalizations $U_q$, $S_q$ and $F_q$ (see below). The
question is whether there exists an  underlying dynamics for
nonextensive thermodynamics. Is it
given by the standard vNE (as worked out in \cite{ppdyn})
or, maybe the dynamics should also be
$q$-modified? 

The analysis given below is based on one single {\it Ansatz\/}: the
$q$-averaged energy $U_q$ is also the Hamiltonian function of the
quantum system. As we shall see, this
implies that $F_q$ is again a stability function for the microscopic
dynamics. In particular, the equilibrium states of
$q$-thermodynamics are dynamically stable.

The vNE is an immediate consequence of the Schr\"odinger
equation (SE)
\be
i|\dot\psi\rangle=H|\psi\rangle\label{S}
\ee
if $\rho$ represents a pure state,
i.e.~$\rho=\varrho=|\psi\rangle\langle\psi|$ (we use $\varrho$
to denote pure state density matrices or their reductions to subsystems).
However, in real experimental situations one does not deal with
pure states since (a) there exists a classical lack of
knowledge about quantum sources and (b) entangled states lead to
non-pure density matrices when reduced to a subsystem. 
Indeed, if the {\it classical\/} state of the
device that prepares the quantum ensemble is not exactly known,
and the experiment is repeated several times,
one has to apply a purely classical averaging over the
classical configurations of the source. This results in a
non-pure density matrix
\be
\rho=\int d\varrho\,w(\varrho)\varrho
\label{edm}
\ee
where one integrates over quantum-mechanical pure states
whose distribution is given by $w(\varrho)$. $\rho$
can be used to calculate experimental averages since the latter
are linear in $\varrho$:
\be
\Tr \rho A=\int d\varrho\,w(\varrho)\Tr\varrho A.
\ee
The distinction between the {\it state\/} $\varrho$ and the distribution
$w(\varrho)$ (function defined on the space of states)
leads ultimately to a sharp distinction
between the Liouville equation (LE) and the vNE. In
order to explain this main point let us first note
that the vNE (\ref {vN}) can be written as a classical Hamiltonian system
\cite{HL,Bona,DW,Jordan,Mor}:
\be
i\dot\rho_{a}=\{\rho_{a},\langle H\rangle_1\}\label{LPvN}
\ee
where $\rho_{a}=\langle \alpha|\rho|\alpha'\rangle$ are the components
of $\rho$ taken in some basis, $\langle H\rangle_1=\Tr \rho H$ is
the average energy, and the bracket
\be
\{A,B\}=
\rho_a\Omega{^{a}}_{bc}\frac{\delta A}{\delta\rho_b}
\frac{\delta B}{\delta\rho_c}.
\label{LPb}
\ee
is a Lie-Poisson bracket on the manifold of states \cite{MR}.
Here $\Omega{^{a}}_{bc}$ are structure constants of a Lie
algebra [$gl(n,C)$ for finite dimensional Hilbert spaces] and
the summation convention means summation or integration with
respect to an appropriate measure \cite{MCpla,MCpra98}.
Using the standard argument one derives the LE
\be
i\dot w=\{w,\langle H\rangle_1\}.\label{L}
\ee
It is obvious that the linearity in $w$ 
of the LE (\ref{L}) 
is completely unrelated to the linearity of the vNE,
and follows directly from the fact that
the dynamics (\ref{LPvN}) is Hamiltonian.
Also the physical meaning of the linearity of (\ref{L}) is
clear: It reflects the {\it linearity of averaging\/} and the
fact that an experimentalist can control the form of $w$
by improving the measurement device.

The second class of density matrices that are non-pure,
in the sense that $\varrho^2\neq \varrho$,  occurs as an entirely
quantum phenomenon and is a result of entanglement between
correlated quantum systems.
Such states are fundamentally and
irreducibly mixed. They can be written in
different ways as convex combinations of pure states and all
such combinations have to be regarded as physically equivalent
\cite{Mielnik,Mermin}.

On the other hand, the decomposition (\ref{edm}) is uniquely
determined by the experimental setup.
The difference in physical status of the ``pure-state" decompositions of
$\rho$ and $\varrho$ implies that 
there exists a 
physical difference between the linearity of the LE (\ref{L}) and this of
the vNE (\ref{vN}).  
The linearity of the latter is a {\it postulate\/}
that is independent of both the linearity of (\ref{L}) and this of the
pure state SE (\ref{S}). Linear SE is compatible with any equation of the
form
\be
 i\dot\varrho=[H,f(\varrho)]\label{fvN}
\ee
provided $f(\varrho)=\varrho$ for
$\varrho^2=\varrho$, which holds for all functions satisfying $f(0)=0$ and
$f(1)=1$. The choice of $f(x)=x$ is convenient but does not seem to be
dictated by any fundamental principle. On the contrary, we will argue that
other choices of $f$ may be physically relevant and, in particular, we
will show that there exists a link of $f(x)=x^q$ with the nonextensive
$q$-statistics introduced by Tsallis \cite{Tsallis} [note that $q>0$ is
needed to ensure $f(0)=0$].

We shall first show that the modified dynamics given by
Eq.~(\ref{fvN}) has the same Lie-Poisson structure as (\ref{vN}).
Assume $f$ has a Taylor expansion
$f(x)=\sum_{k=1}^\infty f_kx^k$
with radius of convergence at least 1.
Consider the Lie-Poisson dynamics
\be
i\dot\varrho_a=\{\varrho_a,\langle H\rangle_f\}\label{fLPvN}
\ee
with the 1-homogeneous Hamiltonian function
\be
\langle H\rangle_f
=\Tr\left\{ (\Tr \varrho)f\left({\varrho\over\Tr \varrho}\right)H\right\}.
\ee
Variation of $\langle H\rangle_f$ with respect to $\varrho$
gives the effective Hamiltonian
\be
{}&{}&\hat H(\varrho)
={\delta \langle H\rangle_f\over\delta\varrho}=
\sum_{k=1}^\infty f_k(\Tr\varrho)^{1-k}
\sum_{m=0}^{k-1}\varrho^{k-1-m}H\varrho^m
\nonumber\\
&{}&\pp =
+\Tr\left\{ f\left({\varrho\over\Tr\varrho}\right)H\right\}\bbox 1
-\Tr\left\{{\varrho\over\Tr\varrho}f'\left({\varrho\over\Tr\varrho}\right)H\right\}\bbox
1,\nonumber
\ee
where $f'=df/dx$.
(\ref{fLPvN}) when written in an operator notation is
\be
i\dot\varrho=[\hat H(\varrho),\varrho]=
(\Tr\varrho)\big[H,f(\varrho/\Tr\varrho)\big].
\ee
$\Tr\varrho$ is an integral of motion so we can consider solutions
normalized by $\Tr \varrho=1$ which shows that (\ref{fvN}) is
indeed a particular case of (\ref{fLPvN}). 1-homogeneity of
$\langle H\rangle_f$ implies that
$
\langle H\rangle_f=\Tr\varrho\hat H(\varrho).
$
Taking $f(x)=x^q$, a normalized $\varrho$, and denoting the
corresponding $\langle H\rangle_f$ by $\langle H\rangle_q$, we can see that
$
\langle H\rangle_q=\Tr\varrho\hat H(\varrho)=\Tr\varrho^qH
$
i.e. the average effective energy equals the
$q$-average of $H$, an internal energy typical of Tsallis
generalized thermodynamics. Such averages were shown to be naturally
linked to nonextensive $q$-entropies \cite{average,gas,r1,r2,r3,r4,data}.
Obviously, any theory that for some reasons deals with $q$-averages
involves some degree of nonextensivity independently of what kind of
entropy is used.
Let us note that in the above case the
vNE is nonlinear,
\be
i\dot\varrho=[\hat H(\varrho),\varrho]=
\big[H,\varrho^q].\label{qvN}
\ee

Nonlinear vNE has been studied recently in \cite{gri}, and
is used regularly in statistical physics. The nonlinearity
is usually due to friction forces and should be compensated by adding
a noise term to the vNE in order to keep the average energy constant.
However, for a nonlinearity of the form (\ref{fvN}) the energy
is a conserved quantity. Hence there is no need for the
balancing noise term.
vNE's of the form (\ref{qvN}) were independently found
in the context of a Lie-Nambu
dynamics and studied in \cite{MCpla,MCMM}. It was
shown, in particular, that their Hermitean Hilbert-Schmidt solutions possess
time-independent spectra,
an important fact that
allows to treat the solutions $\varrho(t)$ as density matrices.
This implies also that for any $t_1$ and $t_2$ there exists a
unitary transformation satisfying
$
\varrho(t_2)=U(t_2,t_1)\varrho(t_1)U(t_2,t_1)^{-1}.
$
Although $U(t_2,t_1)\neq U(t_2-t_1,0)$ [equality would imply linearity
of evolution] the local generator of $U(t+\epsilon,t)$, for
$\epsilon\to 0$, exists and is our effective
time-dependent Hamiltonian $\hat H\big(\varrho(t)\big)$.
Denote $C_n=\Tr(\varrho^n)$, $n\in \bbox N$. One finds
$
\{C_n,F\}=0
$
for any $F$, 
which shows that $C_n$ are
Casimir
invariants for the dynamics.
The set of invariants contains also all $n$-averages of $H$ since
$
\{\langle H\rangle_n,\langle H\rangle_{m}\}=0
$
for any natural $n$ and $m$.
The dynamics we consider is therefore so regular and so close to
the linear one that one may
wonder whether such equations do possess nontrivial solutions.
Fortunately, the answer is positive. An analytical
Darboux-type technique of solving (\ref{qvN}) for $q=2$
has been recently developed \cite{SLMC} and various explicit
solutions were found. In a classical context the $q=2$ case was discussed
in great detail in relation to Lie-algebraic generalizations of
classical Euler equations (cf. \cite{MR} and
references therein, in particular \cite{Adler,Ratiu}). It is also
well known that similar $q=2$ Lie-Poisson equations describe plasma 
dynamics (in the context of a generalized statistics a
paper of particular relevance is \cite{Boghosian}). 

Since the description we propose is meant to provide a
fundamental quantum background for a generalized statistics it
must be also capable of dealing with collections of nonextensive systems.
This means we have to provide a recipe for extending the von Neumann
dynamics from subsystems to composite systems.
The extension should be
self-consistent in the sense that a dynamics of a subsystem
should be independent of whether the system is considered alone
or as part of a collection of many noninteracting systems.
This is achieved by taking the 2-system Hamiltonian function 
$
\langle H_{I}\rangle^I_{q_1}+
\langle H_{II}\rangle^{II}_{q_2}.
$
The $q_k$-averages occuring in the
Hamiltonian function of the composite system are
\be
& &\langle H_I\rangle^I_{q_1}= \Tr_I\big((\varrho_I)^{q_1} H_I\big),\cr
& &\langle H_{II}\rangle^{II}_{q_2}=
\Tr_{II}\big((\varrho_{II})^{q_2} H_{II}\big)\nonumber
\ee
and the density matrices $\varrho_I$ and $\varrho_{II}$ are the {\it
reduced\/} density matrices of the respective subsystems.
The two-system equation one obtains is
\be
i\dot \varrho_{I+II}=[\hat H_I(\varrho_I)\otimes\bbox 1_{II}+
\bbox 1_I\otimes \hat H_{II}(\varrho_{II}), \varrho_{I+II}],\label{qLN1+2'}
\ee
where $\hat H_{I}(\varrho_{I})$, $\hat H_{II}(\varrho_{II})$ are the
effective Hamiltonians of the subsystems.
The choice
of this particular type of extension follows from the general
results proved for a Lie-Poisson nonlinear quantum mechanics of
density matrices in \cite{Jordan,MCpla,MCpra98,MCMK98}. The expressions
$
C_n(\varrho_{I+II})=
\Tr_{I+II}(\varrho_{I+II})^n
$
are time-independent (as Casimir
invariants) for any natural $n$. Therefore, if $\varrho_{I+II}(t)$
is a Hermitean Hilbert-Schmidt solution of (\ref{qLN1+2'}) then
its eigenvalues are time independent on the basis of the
standard argument
\cite{MCMM}, and therefore $\varrho_{I+II}(t)$
is a density matrix if it is one at $t=0$. On the other hand,
taking partial traces of (\ref{qLN1+2'}) one
verifies that
\be
i\dot \varrho_{I}= [H_I,(\varrho_I)^{q_1}],\quad
i\dot \varrho_{II}= [H_{II},(\varrho_{II})^{q_2}],\nonumber
\ee
as required by the self-consistency of the extension. All these
results have an immediate extension to more general Hamiltonian
functions $\langle H\rangle_f$.

Having established all these general results we are now in
position to discuss in more detail the links to the generalized
statistics proposed by Tsallis \cite{Tsallis}. It is based on the 
internal energy 
$U_q=\Tr\varrho^qH$ and the corresponding
entropy 
\be
S_q(\varrho)=k_B\frac{\Tr\varrho-\Tr (\varrho^q)}{q-1}
\label{tsen}
\ee
$U_q$ is naturally
associated with $S_q$ since then various relations typical of
$q=1$ thermodynamics turn out to be $q$-independent. 
However, standard thermodynamics is {\it static\/} and
the relations between $U_q$ and $S_q$ are evaluated in thermal
equilibrium. From the dynamical point of view an
equilibrium state $\varrho_0$ is a fixed point of the dynamics, i.e.~ 
$[H,\varrho_0]=0$. There exists an infinite number of such
states but not all of them have to be stable if a nonlinear
Lie-Poisson dynamics is involved. The stability tests that are typically
used in such a situation (say, in hydrodynamics and plasma physics) are the
energy-momentum, energy-Casimir \cite{Phys.Rep.} 
or energy-invariants \cite{J-P}
methods. In the energy-Casimir method (used when one knows the Casimirs 
but does not control the symmetries) one looks for minima or maxima of
the ``stability function" 
\be
X(\varrho)=h(\varrho)+\Phi(C_1,C_2,\dots)
\ee
where $h(\varrho)$ is a Hamiltonian function of the Hamiltonian
dynamical system and $\Phi$ is a function of the Casimir
invariants $C_k$ typical of this system. The latter function is
determined by the requirement that 
$X(\varrho)$ has a strict minimum or maximum at $\varrho_0$,
in particular
\be
\frac{\delta X}{\delta\varrho}(\varrho_0)=0.
\ee
In our case $h(\varrho)$
is $U_f=\langle H\rangle_f$ (or $U_q$ if we restrict
the analysis to $f(x)=x^q$), and the Casimirs are all functions 
$C(\varrho)$ that can be written as a trace of a convergent power
series i.e.
\be
C(\varrho)=\Tr\big(\sum_kc_k\varrho^k\big)=\sum_kc_kC_k(\varrho)=:
\Phi(C_1,C_2,\dots).\nonumber
\ee
It is clear that the stability
function $X$ for the energy-Casimir method is nothing else but the
free energy $F$ corresponding to a generalized entropy $S=-\Phi/T$
with $T$ the temperature. In
this way the thermodynamic relation $F=U_f-TS$ is recovered.
The equilibrium state $\varrho_0$ is an extremum of $F$. If this is a
strict minimum (or maximum) then the orbits of density matrices
in a neighborhood of $\varrho_0$ are dynamically stable.
Thermodynamic stability of the Tsallis thermodynamics has
been raised in \cite{stab1} and settled in \cite{stab2}.
Therefore equilibrium states extremizing $F$ will generically be
dynamically stable fixed points of the nonlinear vNE.

Once dynamic stability of $\varrho_0$ is established it becomes
meaningful to study linear response theory \cite {KR}. This has been
done in the context of non-extensive statistics by
Rajagopal\cite{r3,raja2}. However, the theory has to be modified because of
the nonlinearity of the vNE. A discussion of these modifications is
out of the scope of the present paper and will be presented
elsewhere.

Let us illustrate the above results with the simple
example of a single spin in an external field. The
Hamiltonian is given by $H=-\mu\sigma_z$ (assume $\mu >0$,
the $\sigma_\alpha,\alpha=x,y,z$ are the Pauli matrices).
A general Hermitean $2\times 2$ matrix with eigenvalues
$\lambda_1,\lambda_2$ is
\be
\varrho&=&{1\over 2}(\lambda_1+\lambda_2){\bbox 1}
+{1\over 2}(\lambda_1-\lambda_2)\cos(\phi)\sigma_z\cr
&-&{1\over 2}(\lambda_1-\lambda_2)\sin(\phi)
\left(\cos(\psi)\sigma_x+\sin(\psi)\sigma_y\right).
\label{gst}
\ee
If $\sin(\phi)=0$ or $\lambda_1=\lambda_2$
then $\varrho$ is time-invariant. In the other case
Eq.~(\ref{fvN}) implies $\dot\phi=0$,
\be
\dot\psi=-\omega
\hbox{ with }
\omega=2\mu{f(\lambda_1)-f(\lambda_2)\over\lambda_1-\lambda_2}
\label{omega}
\ee
and the Larmor precession frequency $\omega$
depends in general on the eigenvalues of $\varrho$.
For this example this is
the only effect induced by the nonlinearity of the vNE.
Note that $\omega=2\mu$ is still valid for pure states
--- the underlying quantum mechanics remains unchanged.

One verifies that
$
\langle H\rangle_q=-\mu\left(\lambda^q-(1-\lambda)^q\right)
$
and
$
\hat H(\varrho_0)=-\gamma\sigma_z
+\hbox{ a multiple of }{\bbox 1}
$
with
$
\gamma=\mu q\big[\lambda^q+(1-\lambda)^q\big]/2.
$
Now assume $f(x)=x^q$.
The internal energy is
\be
U_q=\Tr\varrho^q H=-\mu\cos(\phi)\big(\lambda_1^q-\lambda_2^q\big).
\ee
The $q$-entropy equals
$
S_q=k_B\big[1-\lambda_1^q-\lambda_2^q\big]/(q-1)
$
In both formulas one should take
$\lambda_1=\lambda\ge 0$ and $\lambda_2=1-\lambda\ge 0$.
Of physical interest are minima of $F$ for which $\lambda>1-\lambda$
and $U_q<0$. Hence one can take $\cos(\phi)=1$.
From $\partial F/\partial\lambda=0$,
with $\beta=1/k_BT$ and assuming
$
0<|q-1|\beta\mu< 1
$
one finds that the eigenvalue $\lambda$ of $\varrho_0$ is the solution of
\be
\left({\lambda\over 1-\lambda}\right)^{q-1}
={1+(q-1)\beta\mu\over 1-(q-1)\beta\mu}.
\ee
The value of
$\partial^2F/\partial\lambda^2$
at equilibrium is strictly positive. In fact, $F$ has an absolute
minimum at $\varrho=\varrho_0$. This thermodynamic stability implies
dynamic stability of $\varrho_0$ as a fixed point of the nonlinear vNE.
One concludes that Tsallis thermodynamics is useful to
analyse the dynamic stability of fixed points of the nonlinear vNE.

A special feature of the nonlinear vNE is that classical mixtures
of initial conditions evolve irreversibly with time
(this property has been discussed extensively in \cite{gri}).
Take e.g.
\be
\rho
&=&\int_0^1\hbox{ d}\lambda
\int_0^\pi\sin(\phi)\hbox{ d}\phi\int_0^{2\pi}\hbox{ d}\psi
\ w(\lambda,\phi,\psi)\varrho(\lambda,\phi,\psi)
\nonumber
\ee
with
$w(\lambda,\phi,\psi)=(1/8)\sin(\psi/2)$
and with $\varrho(\lambda,\phi,\psi)$ given by (\ref{gst}).
A short calculation then shows that
\be
\rho(t)&=&{1\over 2}{\bbox 1}
+{\pi\over 24}\int_0^1d\lambda
(2\lambda - 1) \nonumber\\
&\pp =&
\pp{
{1\over 2}{\bbox 1}
+{\pi\over 24}}
\times
\Big(\cos(\omega(\lambda)t)\sigma_x
+\sin(\omega\big(\lambda)t\big)\sigma_y\Big)
\nonumber
\ee
with $\omega(\lambda)$ given by (\ref{omega}).
Due to the dependence of $\omega$ on $\lambda$ a dephasing
occurs and the classical density matrix converges
to $\frac{1}{2}{\bbox 1}$ as $t\rightarrow\infty$.
The lack of knowledge about initial conditions leads to a true
irreversible decay.

An analysis of more complicated examples including linear
response theory will be presented in a forthcoming paper.

We are grateful to Juan-Pablo Ortega for his comments on stability of
nonlinear Hamiltonian systems and for his interest in our work. 
Our collaboration is a part of the Flemish-Polish project 
No.~007 and was financed in part by the KBN Grant No.
2~P03B~163~15.


\begin{references}
\bibitem{Mermin}N.~D.~Mermin, e-print quant-ph/9801057; this is
the most detailed and clear presentation of the problem we have
found in the literature.
\end{references}
\end{document}